\documentclass[12pt]{article}
\usepackage{graphicx}

\begin{document}

\begin{titlepage}

\begin{center}

\vspace*{5cm}

{\Large {\bf Self-annihilation of the neutralino dark matter into
two photons or a $Z$ and a photon in the MSSM.}}

\vspace{8mm}

{\large F. Boudjema${}^{1)}$, A.~Semenov${}^{2)}$ and D.~Temes${}^{1)}$ }\\

\vspace{4mm}

{\it 1) LAPTH$^\dagger$, B.P.110, Annecy-le-Vieux F-74941, France}
\\ {\it
2) Joint Institute of Nuclear Research, JINR, 141980 Dubna, Russia }\\

\vspace{10mm}

\today
\end{center}

\centerline{ {\bf Abstract} } \baselineskip=14pt \noindent

{\small We revisit the one-loop calculation of the annihilation of
a pair of the  lightest neutralinos into a pair of photons, a pair
of gluons and also a $Z$ photon final state. For the latter we
have identified a new contribution that may not always be
negligible. For all three processes we have conducted a tuned
comparison with previous calculations for some characteristic
scenarios. The approach to the very heavy higgsino and wino is
studied and we argue how the full one-loop calculation should be
matched into a more complete treatment that was presented recently
for these extreme regimes. We also give a short description of the
code that we exploited for the automatic calculation of one-loop
cross sections in the MSSM that could apply both for observables
at the colliders and for astrophysics or relic density
calculations. In particular the automatic treatment of zero Gram
determinants which appear in the latter applications is outlined.
We also point out how generalised non-linear gauge fixing
constraints can be exploited.}

\vspace*{\fill}

\vspace*{0.1cm} \rightline{LAPTH-1106/05}

\vspace*{1cm}

$^\dagger${\small UMR 5108 du CNRS, associ\'ee  \`a
l'Universit\'e de Savoie.} \normalsize

\vspace*{2cm}

\end{titlepage}


\def\mww{m_w^2}
\def\sww{s_w^2}
\def\cww{c_w^2}
\def\beqn{\begin{eqnarray}}
\def\eeqn{\end{eqnarray}}
\def\mneuto{M_{\tilde{\chi}^{0}_{1}}}
\def\tgb{\tan\beta}
\def\tb{\tan\beta}
\def\ra{\rightarrow}
\def\neuto{\tilde{\chi}^0_1}
\def\nngg{$\neuto \neuto \ra \gamma \gamma \;$}
\def\nnzg{$\neuto \neuto \ra Z \gamma \;$}
\def\nnzgt{$\neuto \neuto \ra Z \gamma $}
\def\nnglgl{$\neuto \neuto \ra g g \;$}
\def\noi{\noindent}





\renewcommand{\topfraction}{0.85}
\renewcommand{\textfraction}{0.1}
\renewcommand{\floatpagefraction}{0.75}

\section{Introduction}
We now have overwhelming evidence that ordinary matter accounts
for a minute portion of what constitutes the Universe at large.
Most impressive is the confirmation from the very recent WMAP
data\cite{wmap}. Very interestingly,  many extensions of the
standard model whose primary aim was related to the Higgs sector
and the mechanism of symmetry breaking do provide a good dark
matter candidate. Very soon, with the energy frontier that will
open up at the  LHC, intense searches for this new physics with
its
associated dark matter candidate will be pursued in earnest.\\
\noindent Meanwhile, many astroparticle experiments are going on,
and will be improved by the time the LHC runs, to detect dark
matter particles. The problem, either for direct or indirect
detection of dark matter outside the colliders, is that we do not
control many astrophysical parameters. For indirect detection
which is the result of the annihilation of a dark matter pair in,
say, the galactic halo of our galaxy, the photon signal is cleaner
than that of the charged positron and antiproton that are
considered as sources of exotic cosmic rays. The photon will point
back to the source while the antiproton flux suffers from
uncertainties due to the propagation. Of course in both cases one
still has to rely on a modelling of the dark matter profile since
one needs to know the number density of the annihilating dark
matter particles. A very distinctive signal though would be that
of a ``direct" annihilation into a monochromatic photon. In this
case the spectrum will reveal a peak at an energy corresponding to
the mass of the annihilating particles since the latter move at
essentially zero relative velocity $v$. In the galactic halo
$v/c\sim 10^{-3}$. Therefore, provided one has a detector with
good energy resolution, the flux from the ``direct" annihilation
will clearly stand out above the (astrophysical) background or the
diffuse contribution. The latter is due essentially to
annihilation into quarks and $W$ which subsequently fragment and
radiate/decay into photons. This contribution has a continuous
featureless energy distribution which is only cut-off at a maximum
energy corresponding to the mass of the dark matter particle.
There are, and there will be, many powerful detectors to search
for such photon signals, covering a wide range in energy from MeV
to TeV. These are either ground based, like the atmospheric
Cerenkov telescopes, ACT,({\tt Cangaroo}\cite{cangaroo}, {\tt
HESS}\cite{hess}, {\tt MAGIC}\cite{magic}, {\tt
VERITAS}\cite{veritas},..) or space borne telescopes ({\tt
EGRET}\cite{egret}, {\tt AMS}\cite{ams}, the upcoming {\tt
GLAST}\cite{glast},..). Many see in some of the present data an
excess that might be a sign for New Physics and dark matter
annihilation but we should probably be cautious and await
confirmation from other more precise detectors covering the same
energy range. One should also improve on the theoretical
predictions and a better understanding of the background and the
astrophysical component
that enter the calculation of the photon yield.\\
\noindent Our aim in this paper is to revisit the calculation of
the ``direct"  self-annihilation into $\gamma \gamma$, $Z\gamma$
and $gg$ of the lightest supersymmetric particle (LSP). This is  a
neutralino that we will  denote as $\neuto$. There have been a few
attempts of calculating the one-loop induced $\neuto \neuto \ra
\gamma \gamma$ before two complete calculations\cite{lsptogg}
settled the issue. These calculations have been made in the limit
$v=0$ as is appropriate for dark matter annihilation in the halo.
A very recent calculation\cite{lsptoggfernand} has also been made
for this mode. Their results for $v=0$ agree in their most
important features (higgsino limit, for example) with those of
Refs~\cite{lsptogg}, but as far as we are aware no systematic
comparison has been performed. Much more important however is that
there is, at the moment, only one calculation of
\nnzg\cite{lsptozg} (performed at $v=0$) despite the fact that new
features appear in this computation. These features, as we will
see, can not be a mere generalisation of the $2\gamma$ final
state. We will in this paper calculate both \nngg and \nnzg for
any velocity and make a tuned comparison with the existing codes
for $v=0$, {\tt DarkSUSY}\cite {DarkSUSY} for \nngg and \nnzg and
{\tt PLATONdml} for \nngg.  In \nnzg we have identified a new
contribution not taken into account in\cite{lsptozg}. We will also
show some
results for $v=0.5$ for both processes. \\
As a by-product we will also compute the self-annihilation into
gluons:
\nnglgl\cite{manuel-neutneutglueglue}. This can be derived from
\nngg by only keeping the coloured particles and dealing properly
with the colour structure. This process could contribute to, for
example, the antiproton signal. We will see that our results for
\nnglgl completely agree with  those of {\tt
DarkSUSY}\cite{DarkSUSY}  and {\tt PLATONdmg}\cite{lsptoggfernand}
for $v=0$ and with {\tt PLATONgrel}\cite{lsptoggfernand} for
$v=0.5$.

\def\muo{{\tt micrOMEGAs}}
\noindent As is known\cite{lsptogg,lsptozg}, the largest contributions to
\nngg and \nnzg, especially for large
neutralino masses, occur when the neutralino is a wino or a
higgsino. As first pointed out in \cite{lsptogg,lsptozg} the cross
section times the relative velocity, $\sigma v$, for both modes,
tends to an asymptotic constant value that scales as $1/M_W^2$ for
$v=0$ and large LSP mass. This result which breaks unitarity is
due to the one-loop treatment of a ``threshold" singularity that
is nonetheless regulated by $M_W$. It has very recently been,
admirably, shown\cite{lspgg-nojiri} how to include the higher
order corrections through a non-relativistic non-perturbative
approach. The latter reveals the formation of bound states with
zero binding energy that show up as sharp resonances that
dramatically enhance the cross section for particular masses. We
have therefore thought it worthwhile to study the one-loop
derivation in these scenarios and see how one can match the
non-perturbative regime. The reason we do this and the main reason
we carry the calculation of \nngg and \nnzg is that one needs a
reliable code for the photon flux from self-annihilating
neutralino LSP's. These cross sections have been missing from {\tt
micrOMEGAs}\cite{micromegas-all} that we have been developing for
a very accurate derivation of the relic density in supersymmetry
and which is currently being adapted to direct and indirect
detection. The present paper only deals with the cross section
calculation, leaving aside the astrophysical issues to the
implementation and exploitation within \muo. See however
Ref.~\cite{brungamma} for a preliminary use of \muo$\;$ to
indirect detection using some of the results of this paper. \\
The results presented in this paper constitute some of the first
applications of a code for the automatic computation of one-loop
processes in supersymmetry relevant both for the colliders and
astrophysics, such as the problem at hand. Most crucial for the
latter is a careful treatment of the loop integrals since for
these applications the use of general libraries is not appropriate
leading to division by zero because of the appearance of vanishing
Gram determinants in the reduction of the tensor integrals. We
will show how to easily circumvent this problem.

\section{Set-up of the automatic calculation}
Even in the standard model, one-loop calculations of $2\ra 2$
processes involve hundreds of diagrams and a hand calculation is
practically impracticable. Efficient automatic codes for any
generic $2\ra 2$ processes, that have now been exploited for many
$2\ra 3$\cite{grace2to3,other2to3} and even some $2\ra
4$\cite{grace2to4,Dennereeto4f} processes, are almost unavoidable
for such calculations. For the electroweak theory these are the
{\tt GRACE-loop}\cite{nlgfatpaper} code and the package {\tt
FormCalc}\cite{FormCalc} based on  {\tt FeynArts}\cite{FeynArts}
and {\tt LoopTools}\cite{looptools}. \\
\noindent With its much larger particle
content, far greater number of parameters and more complex
structure, the need for an automatic code at one-loop for the
minimal supersymmetric standard model is even more of a must. A
few parts that are needed for such a code  have been developed
based on the package {\tt FeynArtsusy}\cite{FeynArtsusy} but, as
far as we know, no complete code exists or is, at least publicly,
available. {\tt Grace-susy}\cite{Grace-susy} is now also being
developed at one-loop and many results
exist\cite{Grace-susy-1loop}. One of the main difficulties that
has to be tackled is the implementation of the model file, since
this requires that one enters the thousands of vertices that
define the Feynman rules. On the theory side a proper
renormalisation scheme needs to be set up, which then means
extending many of these rules to include counterterms. When this
is done one can just use, or hope to use, the machinery developed
for the SM, in particular the symbolic manipulation part and most
importantly the loop integral routines and tensor
reduction algorithms.\\
\noindent The calculations that we are reporting here are based on a
new automatic tool that uses and adapts modules, many of which,
but not all, are part of other codes. We will report on this
approach elsewhere. Here we will be brief. In this application we
combine {\tt LANHEP}\cite{lanhep} (originally part of the package
{\tt COMPHEP}\cite{comphep}) with the {\tt FormCalc} package but
with an extended and adapted {\tt LoopTools}. {\tt LANHEP} is a
very powerful routine that {\em automatically} generates all the
sets of  Feynman rules of a given model, the latter being defined
in a simple and compact format very similar to the canonical
coordinate representation. Use of multiplets and the
superpotential is built-in to minimize human error. The ghost
Lagrangian is derived directly from the BRST transformations. The
{\tt LANHEP} module also allows to shift fields and parameters and
thus generates counterterms most efficiently. Understandably the
{\tt LANHEP} output file must be in the format of the model file
of the code it is interfaced with. In the case of {\tt FeynArts}
both the {\it generic} (Lorentz structure) and {\it classes}
(particle content) files had to be given. Moreover because  we use
a non-linear gauge fixing condition\cite{nlgfatpaper}, the
{\tt FeynArts} default {\it generic} file had to be extended. \\
\noindent This brings us to the issue of the gauge-fixing. We use a
generalised non-linear gauge\cite{nlg-generalised} adapted to the
minimal supersymmetric model. The gauge fixing writes
\begin{eqnarray}
{\mathcal L}_{GF} &=& - \frac{1}{\xi_W}|(\partial_\mu - ie
  \tilde \alpha \gamma_\mu - igc_W \tilde \beta
  Z_\mu) W^{\mu \, +}
+ \xi_W \frac{g}{2}(v + \tilde \delta h + \tilde
  \omega H + i \tilde \kappa \chi_3) \chi^+|^2 \nonumber \\
&& - \frac{1}{2\xi_Z} (\partial . Z + \xi_Z \frac{g}{2c_W}(v +
  \tilde \epsilon h + \tilde \gamma H) \chi_3)^2 -
  \frac{1}{2\xi_\gamma} (\partial . \gamma)^2. \nonumber
\end{eqnarray}

\noi $h$ and $H$ are the CP-even physical Higgses, with $h$ denoting
the lightest. $\gamma$ is the photon field and the masses of the
charged and neutral weak bosons are related through $M_W=M_Z c_W$.
The $\chi$'s are the Goldstone fields. The non-linear gauge fixing
parameters are $\tilde
\alpha$, $\tilde \beta$, $\tilde
\delta$, $\tilde \omega$, $\tilde \kappa$, $\tilde \epsilon$ and
$\tilde
\gamma$. The $\xi$ are the usual Feynman parameters. In our
implementation the latter are set to $\xi=1$ not only to avoid
very large expressions due to the ``longitudinal" modes of the
gauge bosons but most importantly so that high rank tensors for
the loop integrals are not needed. Gauge parameter independence
which is a non trivial check on the result of the calculation can
be made through the non-linear gauge fixing terms. In many
instances a particular choice of the non-linear gauge parameter
may prove much more judicious than another. For the case at hand,
$\tilde{\alpha}=1$, preserves $U(1)_{\rm em}$ gauge invariance
which explains the vanishing of the $W^+\chi^-\gamma$ vertex. This
will prove crucial for the calculation of \nnzg.
\\
This brings us to the implementation of the loop integrals and
their use in the most general application to radiative corrections
in SUSY both for the colliders, for indirect detection and
improvement of the relic density calculation beyond tree-level. In
{\tt LoopTools}\cite{looptools} for example, the tensor loop
integrals are reduced recursively to a set of scalar integrals by
essentially  following the Passarino-Veltman
procedure\cite{PassarinoVeltman}. The reduction involves solving a
set of equations that brings in the inverse Gram
determinant\footnote{For a recent overview of the problem with the
Gram determinant, see
\cite{thomasfatpaper}. We will however present, for the $2\ra 2$ processes, a simple
solution.}.
Although for applications to the colliders the latter only
vanishes for exceptional points in phase space, for the indirect
detection calculation of tensor integrals involving annihilating
LSP's with small relative velocity $v$, the Gram determinant is of
order $v^2$. Therefore it vanishes exactly for $v=0$ or can get
extremely small slightly away from this value rendering the
calculation highly instable. In the Appendix we show how we dealt
with this problem in an automatic implementation. In a nut-shell,
we have used a segmentation procedure based on the fact that when
some momenta are dependent like what occurs with $v=0$, a
$N$-point function writes as a sum of $N-1$ point functions. This
also applies to the tensorial structures. This observation is not
new (see for example\cite{lsptogg,lsptozg}) and has been used
mostly in hand calculations. Some aspects of it may be remotely
related to\cite{robin-reduc}. The scheme also  allows an expansion
away from exactly vanishing Gram determinants.

\begin{figure*}[tbhp]
\begin{center}
\vspace*{0.5cm}
\includegraphics[width=\textwidth,height=8cm]{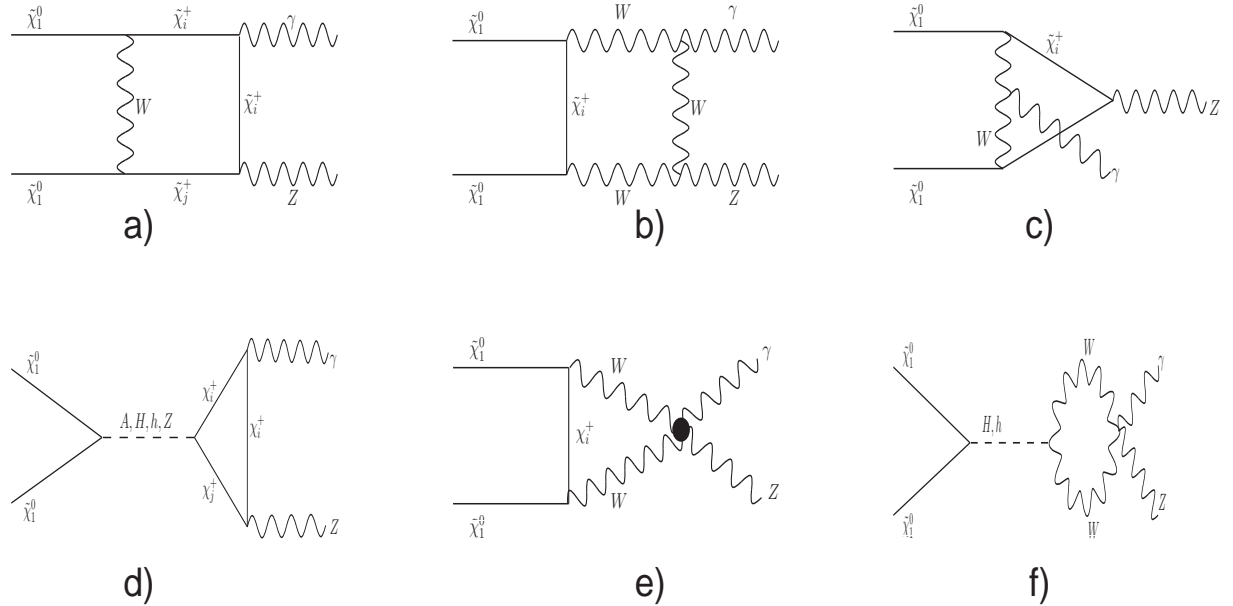}
 \caption{\label{fig.feyn-gg}{\em Typical classes of diagrams
common to both \nngg and \nnzg. For the former, the same
``flavour" circulates in the loops and we do not have any mixing
as in a) and d). Diagrams with Goldstone bosons are not shown. In
the heavy wino and higgsino limit, a) is the dominant diagram.
Diagrams d) and f) with $H,h$ exchange do not contribute for
$v=0$.}}
\end{center}
\end{figure*}

A selection of diagrams that contribute to both \nngg and \nnzg is
shown in Fig.~\ref{fig.feyn-gg} (see also \cite{lsptogg,lsptozg}).
Diagrams of type a) in Fig.~\ref{fig.feyn-gg} are particularly
important in the large wino and higgsino limit. In this limit the
LSP and the internal chargino are of almost equal mass. If the $W$
mass can be neglected this leads to a threshold singularity.\\

\begin{figure*}[htbp]
\begin{center}
\mbox{\includegraphics[width=0.3\textwidth,height=2.5cm]{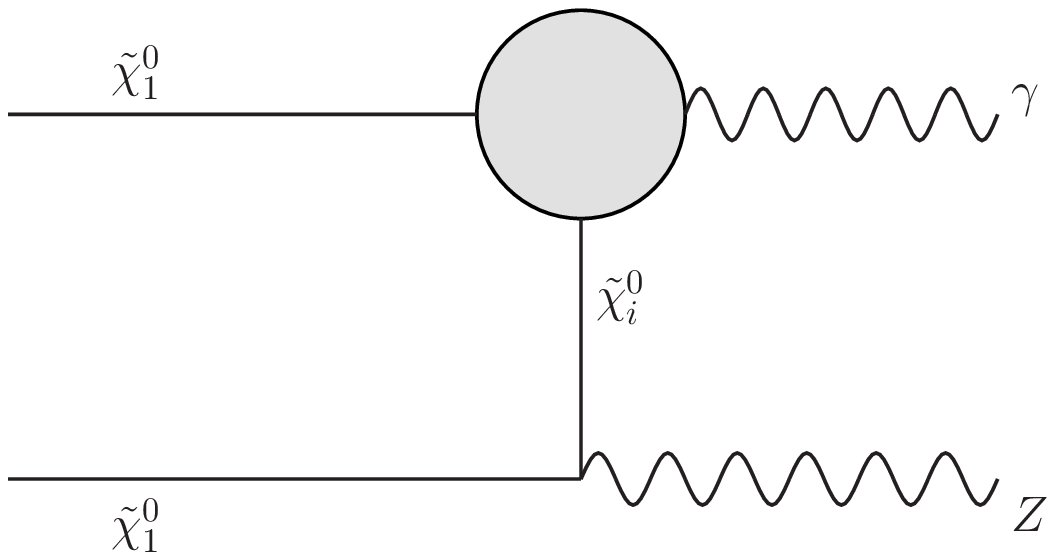}\hspace*{1cm}
\includegraphics[width=0.3\textwidth,height=2.5cm]{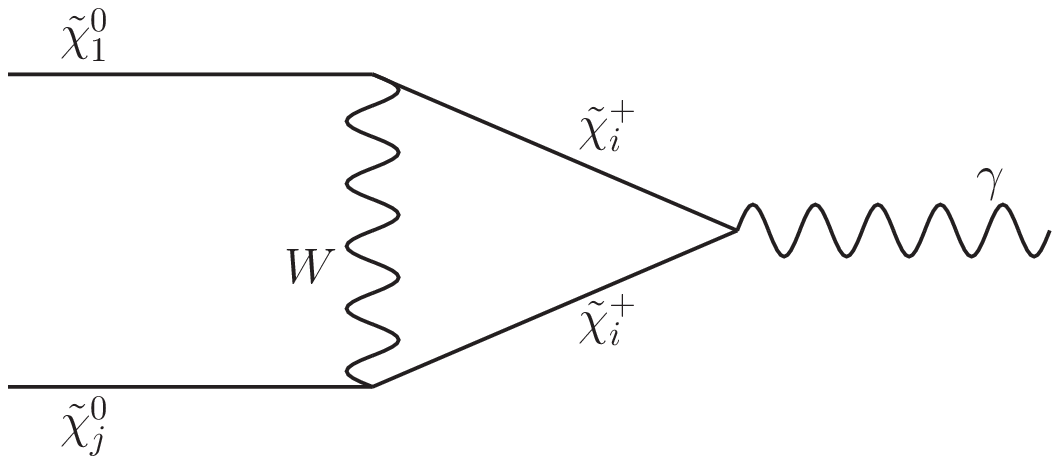}\hspace*{1cm}
\includegraphics[width=0.3\textwidth,height=2.5cm]{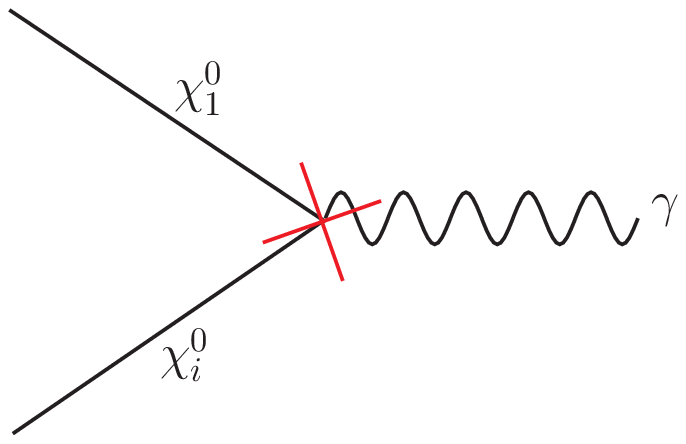}}
\caption{\label{fig.feyn-zg}{\em  An additional class of
diagrams describing the $\tilde{\chi}_i^0 \ra \neuto \gamma$
transition that only appear in the case of \nnzg. A representative
of the blob is the virtual correction. The counterterm
contribution is shown also.}}
\end{center}
\end{figure*}
\noi Moving from \nngg to \nnzg brings mixing effects in the
loops. Most diagrams can be derived from \nngg. There is however
an important class that is only present for \nnzg as shown in
Fig.\ref{fig.feyn-zg}. This class of diagrams is missing
in\cite{lsptozg}. It corresponds to the insertion of the
$\tilde{\chi}^0_i
\ra \neuto \gamma$ transition\footnote{The radiative neutralino
decay $\tilde{\chi}_j^0 \ra \tilde{\chi}_i^0 \gamma$ is calculated
in\cite{HaberWyler}.}. In a general gauge, the virtual transition
would be gauge dependent and not ultraviolet finite. To remedy
both these problems requires the $\tilde{\chi}^0_i \neuto \gamma$
counterterm which is generated starting from the (tree-level)
$\tilde{\chi}^0_i  \neuto Z$ vertex through a $Z-\gamma$ one-loop
transition. It is well known that the latter is gauge dependent,
see for example\cite{nlgfatpaper}. The counterterm requires the
field normalisation $\delta Z_{Z\gamma}^{1/2}$\cite{nlgfatpaper}.
This field renormalisation constant in fact also induces, like in
the standard model, $(H,h)Z\gamma$ vertices not present at
tree-level. This induced vertices are also needed for the class of
diagrams shown in Fig.~\ref{fig.feyn-gg}, in particular those with
(H,h) exchange. The full set of counterterms for \nnzg is in fact
obtained from the tree-level $\neuto \neuto \ra ZZ$, replacing a
$Z$ by a photon and inserting the $\delta Z_{Z\gamma}^{1/2}$
renormalisation constant. However, it is known that  this
renormalisation constant vanishes for
$\tilde{\alpha}=1$\cite{nlgfatpaper}. We have checked this
property explicitly with our code. After this check has been made,
$\tilde{\alpha}=1$ was set, since it considerably reduces the
number of diagrams and most importantly allows to discard all the
counterterm contributions. Further gauge parameter independence of
the result for \nnzg was checked by varying the other non-linear
gauge parameters that enter the calculation, namely
$\tilde{\beta},\tilde{\delta}$ and $\tilde{\omega}$. When
discussing our results for \nnzg we will weight the effect of the
new class  of diagrams shown in Fig.~\ref{fig.feyn-zg} against
that of the full contribution, in doing so we will specialise to
$\tilde{\alpha}=1$. As we will see these diagrams give a non
negligible contribution
especially for the Higgsino case.\\
The application to \nnglgl is rather straightforward. This process
confirms that our code handles the colour summation correctly.
Keeping only one flavour of quark with charge $Q_f$, \nnglgl can
be derived from \nngg through $(N_c Q_f \alpha)^2 \ra 2
\alpha_s^2$ ($N_c=3$ is the number of colours, $\alpha$ is the
electromagnetic fine structure constant and $\alpha_s$ the QCD
equivalent).

\section{Results and comparisons}

We first check that our results are ultraviolet finite by changing
the numerical value of the parameter $1/\epsilon$ that controls a
possible ultraviolet divergence. $\epsilon=4-n$, where $n$ is the
dimensionality of space. We also check for gauge parameter
dependence by varying the non-linear gauge parameters, namely
$\tilde{\alpha},
\tilde{\delta}$ and $\tilde{\omega}$ for \nngg and $\tilde{\beta},
\tilde{\delta}$ and $\tilde{\omega}$ with $\tilde{\alpha}=1$ for \nnzg.
These checks are carried in double precision and show that, for
all points we have studied,  the results are consistent up to $13$
digits. It is important to maintain the relation $M_W=c_W M_Z$. If
these parameters are taken as independent the gauge parameter
independence is lost.
\noi We first discuss our results for $6$ representative scenarios
that we think are good checks on different parts of the
calculations and also because they reveal the most important
characteristics of these cross sections. Moreover these scenarios
also serve to  perform tuned comparisons against codes that are
publicly available. To achieve this it is best to feed the codes
the same parameters. Comparisons that use as input high-scale
values for some SUSY parameters that are run down through some
Renormalisation Group Equation, RGE, package  often need to
specify an interface. Moreover most often the RGE codes are
updated and one does not always have access to the same version to
perform a tuned comparison. For all these scenarios the input
parameters are defined at the electroweak scale and are: $M_1$ the
$U(1)$ gaugino mass,  $M_2$ the $SU(2)$ counterpart, $\mu$ the
Higgsino ``mass", $M_A$ the pseudoscalar mass and $m_{\tilde{f}}$
the common sfermion mass. $\tgb$ is set to $10$. The sfermion
trilinear parameter $A_f$ is set to zero for all sfermions but the
stop, depending on the mass of the latter.  Our Higgs masses here
are tree-level Higgs masses, so we avoided points too close to any
Higgs resonance and the issue of the implementation of the width.
When our code for \nngg,\nnzg and
\nnglgl will be fully incorporated within {\tt micrOMEGAs}, corrected
Higgs masses and mixing angles will be properly implemented in a
gauge invariant manner following\cite{me-3h-2hdm}. This improved
implementation is of relevance only for $v\neq 0$ since the
CP-even Higgses do not contribute when the neutralinos are at
rest. When we refer to cross sections this would in fact refer to
the cross section times the relative velocity $v$, $\sigma v$
expressed in ${\rm cm}^3/s$. In terms of $v$, the total invariant
mass of the neutralinos is
$s=4 m_{\neuto}^2/(1-v^2/4)$.\\
\noi The six scenarios have been chosen so as to represent different
properties, like the gaugino/higgsino content for different masses
of the neutralino. We made no attempt whatsoever to pick up points
that lead to a good relic abundance in accord with {\tt
WMAP}\cite{wmap}. This said, for each point, we give the
corresponding values of the relic density, extracted from {\tt
micrOMEGAs}. One should however keep in mind that different
unconventional histories of the Universe\footnote{A few
possibilities are described in\cite{wmaplhclc-requirements}.}
could alter the usual thermal prediction.

\begin{itemize}
\item \underline{Scenario 1: ``Sugra"}. This reproduces a typical output from
so-called mSUGRA scenarios, although the latter would not produce
a common sfermion mass. The neutralino is mostly bino with mass
around $200$GeV. The lightest chargino is a wino.
\item \underline{Scenario 2: ``nSugra"}. The neutralino is quite
light, about $100$GeV and it is essentially bino. Here the mSugra
relation does not hold, rather $M_2 = 4 M_1$.
\item \underline{Scenario 3: ``higgsino 1"}. The neutralino is a
light higgsino of about $200$GeV. The lightest chargino has a mass
about $6$GeV away.
\item  \underline{Scenario 4: ``higgsino 2"}. The neutralino is a
heavy higgsino of about $4$TeV. It is quite degenerate with the
lightest higgsino-like chargino. The mass diference is about
$0.1$GeV.
\item  \underline{Scenario 5: ``wino 1"}. The neutralino and
lightest chargino are light, about $200$GeV. The mass difference
on the other hand is extremely small $0.01$GeV.
\item \underline{Scenario 6: ``wino 2"}. This is like the previous
example but for TeV masses. The LSP is a wino of mass $4$TeV
completely degenerate with the chargino.
\end{itemize}

\begin{table*}[hbtp]
\begin{center}
\begin{tabular}{|c|c|c|c|c|c|c|}
\cline{2-7}
\multicolumn{1}{c|}{}&  Sugra & nSugra & higgsino-1 & higgsino-2  & wino-1 & wino-2 \\
   \hline
  $M_1$& 0.2 & 0.1 & 0.5 & 20. &  0.5& 20.0   \\
  $M_2$& 0.4& 0.4 & 1.0 & 40. &  0.2&4.0 \\
  $\mu$& 1.0& 1.0 & 0.2& 4.0 &  1.0 &40.0   \\
  $M_A$ & 1.0 & 1.0 &1.0 & 10.  & 1.0 &10.0   \\
  $m_{\tilde{f}}$&0.8  &0.8 &0.8& 10. &0.8 & 10.0    \\
  \hline
$\Omega h^2$& 5.31 &18.8 &$6.41\;10^{-3}$&1.59&$1.16\;10^{-3}$&0.46\\
  \hline \hline
\multicolumn{7}{|c|}{}\\
\multicolumn{7}{|c|}{$\sigma v_{\gamma \gamma} \times 10^{27}$}\\
\hline
v=0&$5.82\; 10^{-5} $&$ 1.58\; 10^{-5}$& $7.01 \;10^{-2} $&$4.71 \;10^{-2} $ &$1.99
$ &$ 1.52$   \\
{\tt PLATONdml}&$5.82\; 10^{-5} $&$ 1.58\; 10^{-5}$& $7.01 \;10^{-2} $&$4.72
\;10^{-2} $  &$1.99 $ &$ 1.53$   \\
{\tt DarkSUSY}&$ 5.81 \; 10^{-5}$&$ 1.58 \; 10^{-5} $& $ 7.02
\;10^{-2}$&$ 4.71 \;10^{-2}$ &$1.99 $ &$ 1.52$
\\ \hline
v=0.5 &$5.94\; 10^{-5} $&$ 1.60\; 10^{-5}$& $1.30 \;10^{-1} $&$5.42 \;10^{-3} $
&$2.36 $ &$ 8.69 \;10^{-2}$   \\
\hline \hline
\multicolumn{7}{|c|}{}\\
\multicolumn{7}{|c|}{$\sigma v_{gg}\times 10^{30}$}\\
\hline
v=0& 2.05& 0.60 &5.74  & 0.33  &19.6  &0.42    \\
{\tt PLATONdmg}& 2.05& 0.60 &5.75  & 0.33  &19.6  &0.42    \\
{\tt DarkSUSY
}& 2.05 & 0.60& 5.77  & 0.33  &19.5  &0.42   \\
\hline
v=0.5 & 2.21&0.60 & 8.23 &0.33  &20.2  &0.42    \\
{\tt PLATONgrel}& 2.21&0.60 & 8.23 &0.33  &20.2  &0.42 \\
\hline \hline
\multicolumn{7}{|c|}{}\\
\multicolumn{7}{|c|}{$\sigma v_{Z \gamma}\times 10^{27}$}\\
\hline v=0,full&$2.03\; 10^{-5} $&$ 2.61\; 10^{-6}$& $2.19
\;10^{-1} $&$2.20
\;10^{-2} $  &$11.7 $ &$ 10.1$ \\
v=0,part &$1.94\; 10^{-5} $&$ 2.50\; 10^{-6}$& $2.61 \;10^{-1}
$&$3.29
\;10^{-2} $ & $11.7 $ &$ 10.1$  \\
{\tt DarkSUSY}&$1.42\; 10^{-5} $&$ 1.79\; 10^{-6}$& $2.61
\;10^{-1} $&$3.29
\;10^{-2} $ & $11.7 $ &$ 10.1$  \\
\hline v=0.5,full& $2.45\; 10^{-5} $&$ 3.67\; 10^{-6}$& $2.99
\;10^{-1} $&$1.66
\;10^{-2} $  &$14.2 $ &$ 5.76\; 10^{-1} $ \\
v=0.5,part& $2.34\; 10^{-5} $&$ 3.53\; 10^{-6}$& $3.58 \;10^{-1}
$&$2.47
\;10^{-1} $ & $14.2 $ &$ 5.76\; 10^{-1} $ \\
 \hline
\end{tabular}
\caption{\label{compar-darksusy}{\em
Results of our calculation both at $v=0$ and $v=0.5$ for \nngg,
\nnzg and \nnzg and comparison with the codes of {\tt PLATON}
and {\tt DarkSUSY}. For \nnzg, ``full" refers to the compete set
of diagrams. ``Part" refers to excluding the $\tilde{\chi}_i^0
\tilde{\chi}_1^0
\gamma$ insertion. Inputs are at the electroweak scale and are expressed in TeV.
$\tgb=10$. $A_f=0$ apart from $A_t=-300GeV$ for
$m_{\tilde{f}_{L,R}}=0.8$TeV and $A_t=0$ for
$m_{\tilde{f}_{L,R}}=10$TeV. We have taken the {\tt DarkSUSY}
inputs, with $M_Z=91.187$GeV and $s_W^2=0.2319$ (but $M_W=M_Z
c_W$). The quark masses are $m_t=175$GeV, $m_b=$5GeV,$m_u=56$MeV,
$m_d=99$MeV, $m_s=199$MeV, $m_c=1.35$GeV.
$\alpha_{em}^{-1}=127.942$ and $\alpha_s=0.117$. The relic
abundance $\Omega h^2$, extracted from {\tt micrOMEGAs}, is also
given for completeness.}}
\end{center}
\end{table*}

Table~\ref{compar-darksusy} shows that the nature of the LSP and
its mass critically determine its self-annihilation cross section
to $\gamma
\gamma$ and $Z\gamma$. The results for the different scenarios
vary by $6$ orders of magnitude, especially for $v=0$. The
bino-like LSP gives far too small cross sections that are unlikely
to be observed as a $\gamma$-ray line. The largest cross sections
\nngg and \nnzg for a LSP mass up to $4$TeV are found for the
wino-like LSP. Moreover in this case the signal is almost an order
of magnitude stronger in $Z\gamma$ than in $\gamma \gamma$,
however the two lines even for $\mneuto \sim 200$GeV are only
$10$GeV away,  even before any smearing is taken into account. For
the wino case the contribution of the $\tilde{\chi}_i^0 \ra \neuto
\gamma$ transition to \nnzg is negligible for the two scenarios we
display in Table~\ref{compar-darksusy}. This is not the case of
the higgsino scenarios (nor the bino-like for that matter) where
this contribution could amount to a correction of more than
$30\%$. It is also interesting to note, see later, that for the
very heavy wino scenario the cross section drops very quickly as
we increase
the velocity. We will study the wino and higgsino case in more detail below.\\
\noi For \nnglgl, the LSP composition does not show dramatic
differences in the cross sections. The largest are however found
for a light wino and a light higgsino of $200$GeV.\\

\noi Let us now turn to the comparisons, essentially for $v=0$, with the codes {\tt
PLATON} and {\tt DarkSUSY}. For this fine-tuned comparison we have
taken the same masses for the quarks and $M_Z$ as in {\tt
DarkSUSY}\cite{DarkSUSY} as well as for the electromagnetic and
strong coupling. On the other hand we imposed $M_W=M_Z c_W$.
Taking for example, $M_W=80.33$GeV, with all other parameters as
in Table~\ref{compar-darksusy} not only gives gauge parameter
dependent results, but
in the wino case the LSP would turn out to be the chargino.\\
\noi Table~\ref{compar-darksusy} shows that our results (for
$v=0$) agree perfectly with those of {\tt PLATONdml} as concerns
\nngg as well as with {\tt PLATONdmg} as concerns \nnglgl. {\tt
PLATONgrel} also perfectly confirms our results for \nnglgl for
$v=0.5$. Excellent agreement with {\tt DarkSUSY} is also observed
(at $v=0$) for \nngg and \nnglgl. To compare with the results of
{\tt DarkSUSY} for \nnzg we do not consider the contribution from
the $\tilde{\chi}_i \ra \neuto \gamma$ ``insertion". With this
restriction we find exactly the same results as {\tt DarkSUSY} in
the case of the wino and higgsino but not in the case of the bino,
where our results are about $30\%$ higher. However as pointed out
earlier, the cross sections in the bino case are
tiny. \\
\noi The effect of the new contribution from $\tilde{\chi}_i \ra
\neuto \gamma$ for both $v=0$ and $v=0.5$ is not noticeable when
the neutralino is a pure wino, but it can be important in the
higgsino case where the total contribution is also large. The new
contribution brings in a relatively small correction in  the
bino case, where the  cross sections  are tiny anyhow.\\
\noi Results for \nngg and \nnzg for $v=0.5$ have been computed here
for the first time and can be relevant for the computation of the
relic density in some regions of the parameter space.\\

\section{The wino and higgsino limits}
\begin{figure*}[tbhp]
\begin{center}
\includegraphics[width=16cm,height=8.5cm]{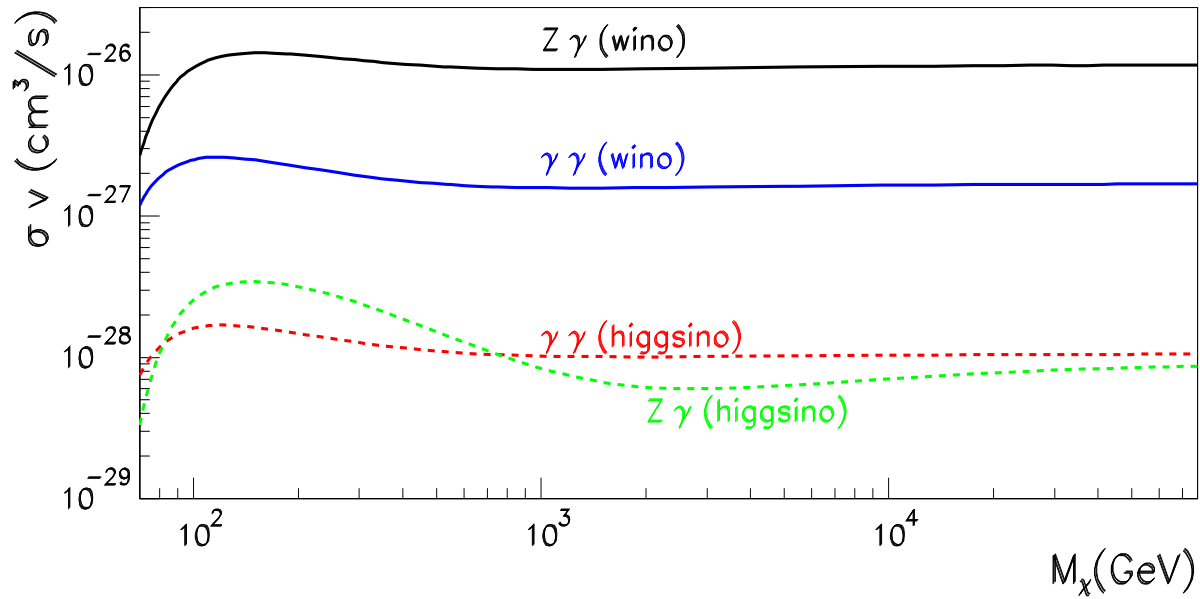}
\caption{\label{fig.plateau-mne}{\em  Dependence of the $\gamma
\gamma$ and $Z \gamma$ cross sections as a function of the LSP
neutralino mass $\mneuto$. For the higgsino case we take $M_2=2
M_1=4\;10^5$TeV and for the wino  $\mu=M_1=2\;10^5$TeV. In both
cases $\tb=10$, $M_A=100$TeV and all sfermions $4\;10^5$TeV. We
use $M_W=M_Z c_W$, with $s_W=0.473$, $M_Z=91.1884$GeV,
$\alpha^{-1}=127.9$.
}}
\end{center}
\end{figure*}

The results of Table~\ref{compar-darksusy} make it clear that most
interesting scenarios for the monochromatic $\gamma$ ray line
signals are of a wino and higgsino type even when the LSP has a
mass of about $2 M_Z$. Fig.~\ref{fig.plateau-mne} shows the
dependence of the
\nngg and \nnzg cross sections at $v=0$ as a function of the LSP
mass in the case of a wino and a higgsino LSP. The mass of the LSP
is in the range $70$GeV to $100$TeV. In fact masses below $100$GeV
may be excluded by LEP2 but it is interesting to see how the cross
sections grow past the $100$GeV mass to stabilise around a
plateau. The masses of the other supersymmetric particles are
taken extremely heavy here. Note that in the higgsino case
\nnzg shows much more structure. The peak cross section is much
more pronounced before the cross section decreases and reaches a
plateau of the same order as \nngg. The wino cross section for
\nnzg is the largest of all and is almost an order of magnitude
larger than \nngg and two orders of magnitude larger than in the
higgsino case.

\begin{figure*}[htbp]
\begin{center}
\includegraphics[width=16cm,height=8.5cm]{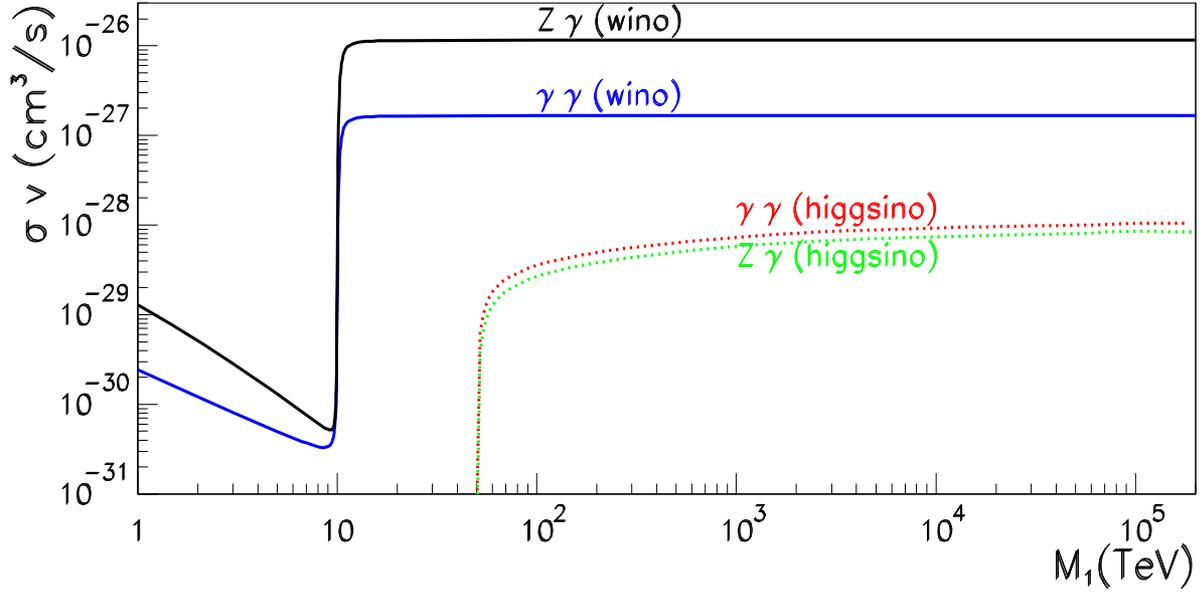}
\includegraphics[width=16cm,height=8.5cm]{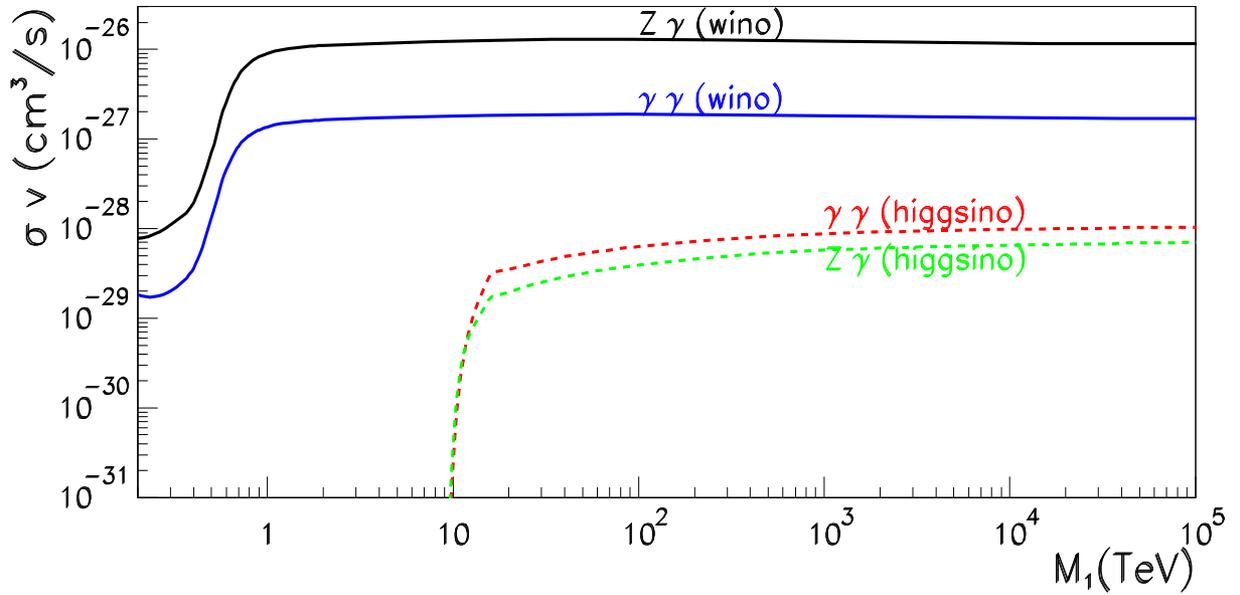}
\caption{\label{fig.plateau-m1}{\em The first figure shows the
dependence on $M_1$ for $\mu (M_2)$ fixed in the higgsino (wino)
limit for $\gamma \gamma$ and $Z \gamma$. The values for the SUSY
parameters are:  $\tgb=10$, $m_{\tilde{f}}= 4\; 10^8$GeV, $A=0$,
$M_A=100$TeV. $\mu=50$TeV, $M_2 = 2 M_1$ in the higgsino case.
$M_2= 10$TeV and $\mu = M_1$ in the wino case. In the figure at
the bottom, the only parameters that are changed are $\mu=10$TeV
in the higgsino case and $\tgb=2$, $M_2=500$GeV in the wino case.
The SM parameters are as in Fig.~\ref{fig.plateau-mne}.}}
\end{center}
\end{figure*}
It is also interesting to see how the plateau is reached for a
fixed mass of the wino and higgsino LSP, or rather a fixed value
of $M_2$ and $\mu$, depending on the composition of the LSP. We
therefore fix these two values and vary the other supersymmetric
parameters of the neutralino sector. The behaviour of the cross
section as we vary these parameters is shown in
Fig.~\ref{fig.plateau-m1}. For the wino case one can see that once
$M_1,\mu > M_2$, and therefore the LSP is mostly wino, the
asymptotic values are reached abruptly especially for the case of
a wino of $10$TeV and large $\tgb$. Below this transition, the
cross sections have a smooth behaviour. In the higgsino limit, a
fast transition occurs once $M_1, M_2 > \mu$ but past this
threshold there is still a smooth and slow increase of the cross
sections before the asymptotic values are reached. \\

Most of this behaviour can, in fact, be recovered through simple
analytical expressions that serve also as a further check on our
results and the accuracy of the calculation in these extreme
scenarios. It had been observed\cite{lsptogg,lsptozg} that when
the LSP is heavy, much heavier that the $W$-boson, the cross
sections (times velocity) for \nngg and \nnzg tend to an
asymptotic value that can be computed from the dominant
contribution, that of diagram a) of Fig.~\ref{fig.feyn-gg}. The
limiting behaviour can be easily understood from the fact that in
the heavy mass limit, the annihilating LSP neutralino and the
chargino are degenerate with a mass much larger than the weak
boson mass. This develops a threshold singularity like what one
finds in QED, although here $M_W$ acts as a regulator.  Another
important factor that measures how the asymptotic values are
reached is the deviation from exact degeneracy between the LSP
neutralino and the lightest chargino given by their mass
difference, $\delta m$, $\delta m = m_{\tilde{\chi}_1^+} -
m_{\tilde{\chi}_1^0}$\cite{lspgg-nojiri}. This scenario has been
revisited in a series of excellent papers\cite{lspgg-nojiri} where
it has been shown how  the one-loop calculation in these cases
need to be improved through a non-relativistic treatment. Our aim
here, in the rest of this section, is to see how our one-loop
results can be made to match with the non-perturbative treatment.
This paves the way to an implementation in a code for indirect
detection that can be used in all generality like what we
have started to do in {\tt micrOMEGAs}. \\
\noi First, we will show how
our one-loop results effectively capture the behaviour of the
cross sections in these scenarios and how the asymptotic value in
the case of a wino is reached dramatically fast. \\
\noi In the higgsino limit, $\mu \ll M_1,M_2$. We
will also take $M_2=2 M_1=2 M_S$ and consider also the large
$\tgb$ (in fact $\tgb>2$ suffices). $\mu$ will be taken
positive.\\
\noi In the wino limit, $M_2
\ll
\mu, M_1$. We will also take $\mu=M_1= M_S$ and large $\tgb$. The
(tree-level) mass difference in the higgsino, $\delta m^{\tilde
h}$, and wino limit, $\delta m^{\tilde w}$, write

\begin{eqnarray}
\delta m^{\tilde h} &\simeq& \frac{m_Z^2}{2M_2}c_W^2 (1-\sin 2\beta) +
\frac{m_Z^2}{2M_1}s_W^2 (1+\sin 2\beta) \sim \frac{5}{16} 
\frac{M_Z^2}{M_S},\nonumber \\
\delta m^{\tilde w} &\simeq& \frac{m_Z^4}{M_1 \mu^2} s_W^2 c_W^2 \sin^2
2\beta\sim  \frac{M_Z^2 M_W^2}{M_S^3} \frac{1}{\tgb^2}.
\end{eqnarray}

\noi We see that in the wino case the mass difference scales like
$1/M_S^3$\cite{lspgg-nojiri}\footnote{It is important to note that
it is essential to have $M_W=M_Z c_W$, otherwise we could get a
mass difference $\propto 1/M_S$.} compared to the $1/M_S$ in the
higgsino case. In these configurations the cross sections are well
approximated\cite{lspgg-nojiri} by $\tilde{\sigma}^{V
\gamma,\tilde{h}}\;v$ in the higgsino case and $\tilde{\sigma}^{V
\gamma,\tilde{w}}\;v$ in the wino case ($V=Z,\gamma$) which are
the results of the dominant diagrams of Fig.~\ref{fig.feyn-gg}-a),

\beqn
\label{asymptotics-reduct}
\tilde{\sigma}^{V \gamma,\tilde{h}}\;v&=&\sigma_\infty^{V \gamma,\tilde{h}} v\left( 1 +
\sqrt{\frac{2m_{\tilde{\chi}_1^0} \delta m}{M_W^2}} \right)^{-2}=\sigma_\infty^{V
\gamma,\tilde{h}}v \left( 1 +
\sqrt{\frac{5 \mu}{6 M_S}} \right)^{-2} , \nonumber
\\
\tilde{\sigma}^{V \gamma,\tilde{w}}\;v&=&\sigma_\infty^{V
\gamma,\tilde{w}} v\left( 1 +
\sqrt{\frac{2 M_Z^2 M_2 }{M_S^3 \tgb^2}} \right)^{-2}; \quad
V=Z,\gamma.
\eeqn

\noi where the asymptotic values ($\delta m =0$) are given by

\beqn
\label{asymptotics-reduc}
\sigma_\infty^{\gamma \gamma,\tilde{h}}\;v&=&\frac{\alpha^4 \pi}{4 M_W^2 s_W^4} \sim
 10^{-28} cm^3/s,\nonumber \\
\sigma_\infty^{\gamma \gamma,\tilde{w}}\;v&=&16 \sigma_\infty^{\gamma
\gamma,\tilde{h}} \;v \sim 1.6 \;10^{-27} cm^3/s, \\
\sigma_\infty^{Z \gamma,\tilde{h}} \;v&=& 2
\frac{(1/2-s_W^2)^2}{s_W^2
c_W^2} \; \sigma_\infty^{\gamma
\gamma,\tilde{h}} \;v \sim  0.8 \; 10^{-28} cm^3/s,\\
\sigma_\infty^{Z \gamma,\tilde{w}} \;v &=&2 \frac{c_W^2}{s_W^2} \;
\sigma_\infty^{\gamma \gamma,\tilde{w}} \;v \sim  10^{-26}
cm^3/s.
\eeqn

We have verified that our code including the complete
contributions agrees extremely well with these approximation for
the cross sections, Eq.~\ref{asymptotics-reduct}, and that
moreover in the wino case the asymptotic values,
Eq.~\ref{asymptotics-reduc}, are reached very fast due to very
small $\delta m$. This is also exemplified in
Fig.~\ref{fig.plateau-m1}.

\begin{figure*}[btp]
\begin{center}
\mbox{\hspace*{-5cm}\includegraphics[width=16cm,height=10.5cm]{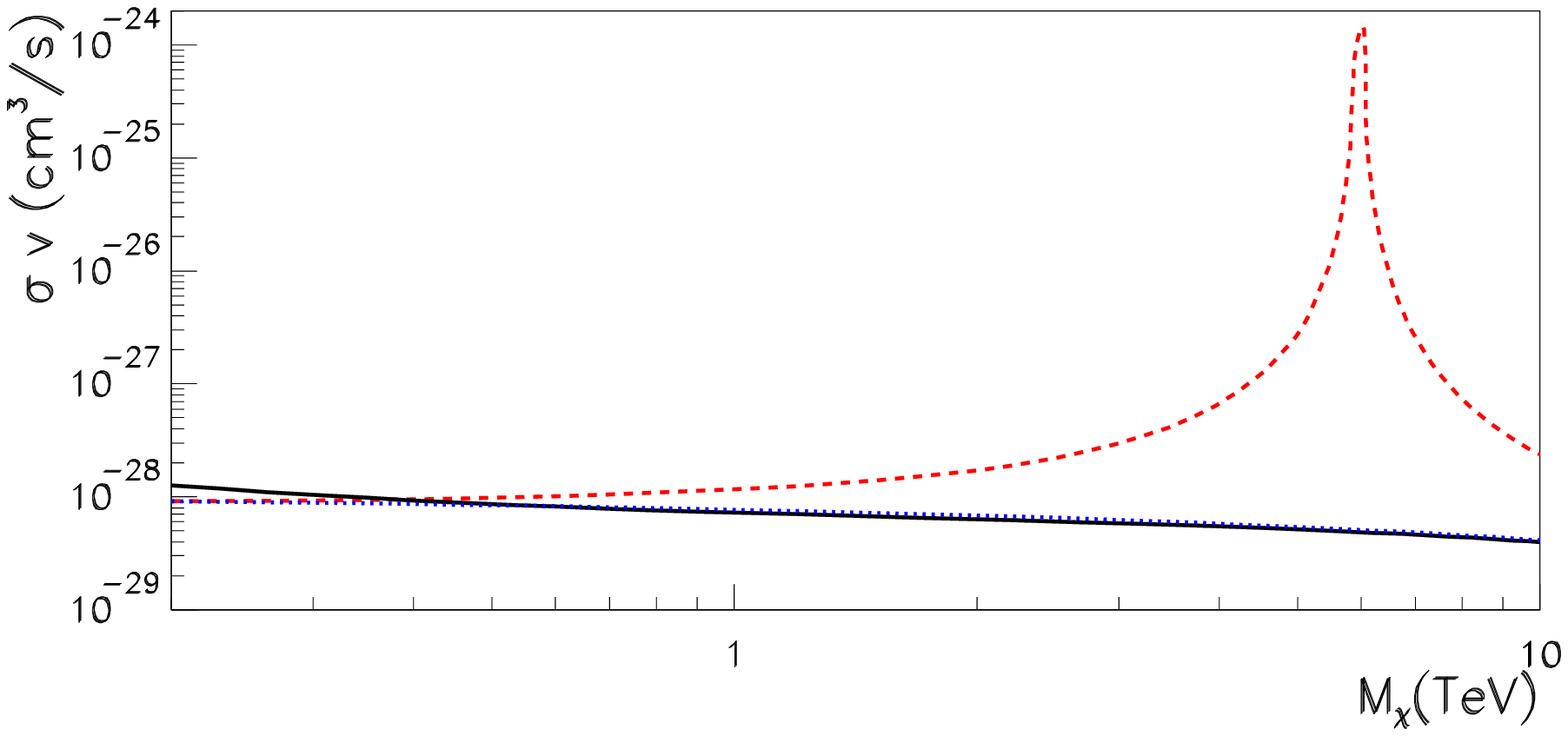}
\raisebox{3.8cm}{ \mbox{ \hspace*{-14.0cm}
\includegraphics[width=8cm,height=5.5cm]{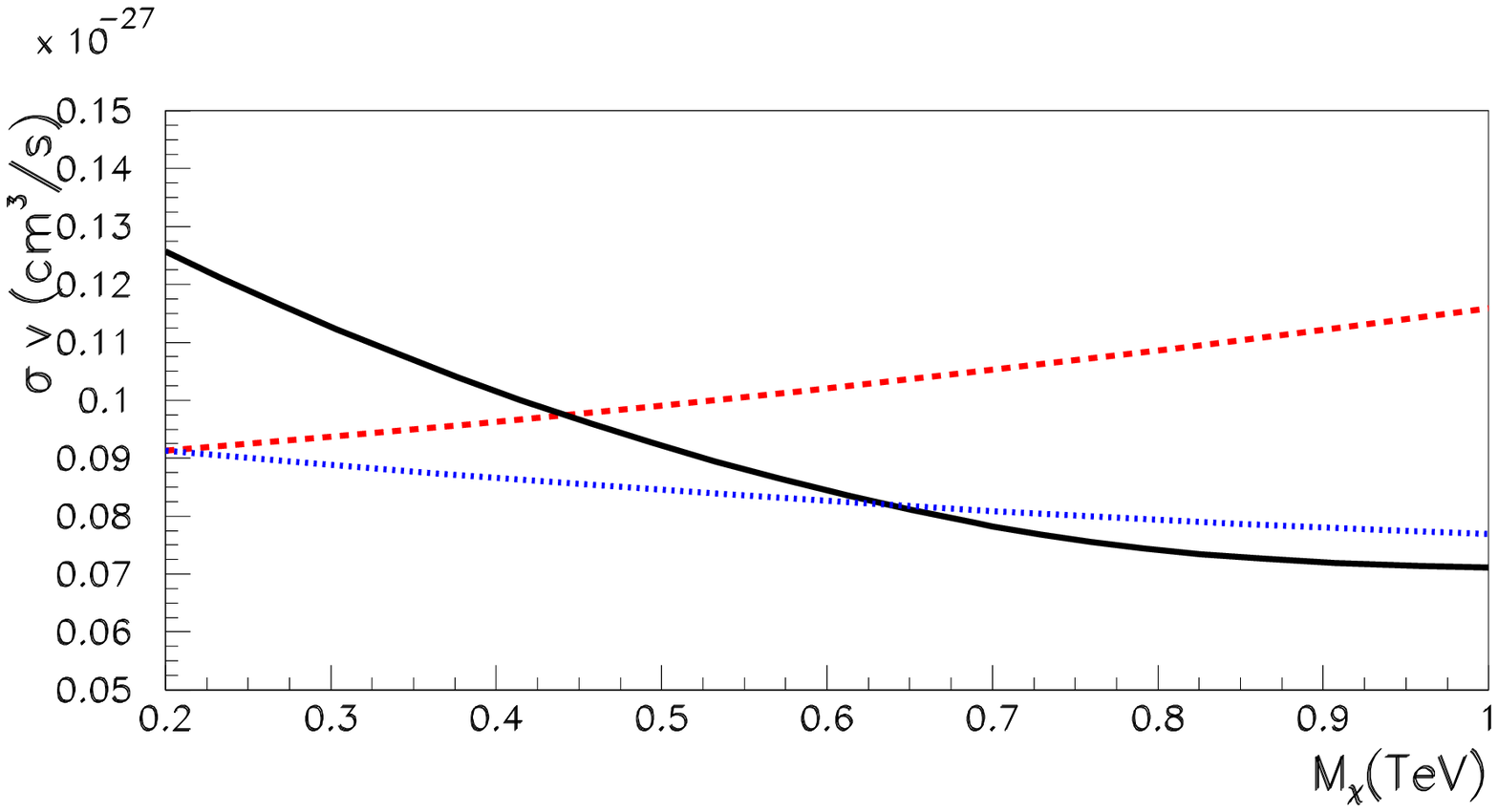}}}}
\caption{\label{fig.plateauM1japon}{\em Comparison, for \nngg as a function
$\mneuto$, between our prediction for the
full one-loop calculation (continuous line, black), analytical
one-loop expressions based on the approximation of
Eq.~\ref{asymptotics-reduct} (dotted, blue) and the
non-perturbative prediction based on the fitting functions
including resonances\cite{lspgg-nojiri} (dashed,red). The input
susy parameters are $M_2=2 M_1=50$TeV, $\tan\beta=10$, $A=0$,
$m_{\tilde f}=M_A=100$TeV. The SM parameters are as in
Fig.~\ref{fig.plateau-mne}. The insert is a close-up for small
$\mneuto$.}}
\end{center}
\end{figure*}

As demonstrated in \cite{lspgg-nojiri} the one-loop treatment of
the threshold singularity that is responsible for the behaviour of
these cross sections in the higgsino and wino regime at high LSP
mass is not adequate an breaks unitarity. The non-perturbative
non-relativistic approach of Ref.~\cite{lspgg-nojiri} not only
improves on the calculation but it also unravels the formation of
bound states that drastically enhance the annihilation cross
sections for specific combinations of masses.
Fig.~\ref{fig.plateauM1japon} shows the effect of such resonances
and the departure of the cross section
\nngg from a full one-loop treatment as the mass of the higgsino LSP
increases. The ``resonance" curve is based on the use of a fitting
function as given in\cite{lspgg-nojiri} \footnote{We  thank
J.~Hisano and M.~Nojiri for confirming that Eq.~61 of
hep-ph/0412403 should be squared and that the entry $i=0,j=1$ in
Table~1 of that paper is 10 times smaller.}. All other particles
are taken heavy apart from the higgsino mass parameter $\mu$ and
$M_1=M_S=M_2/2=25$TeV. For the whole range of $\mneuto \sim \mu$
we have $\delta m=0.1$GeV. The figure also shows the value of the
approximate one-loop result as given for the higgsino limit in
Eq.~\ref{asymptotics-reduct}, together with the full one-loop
treatment based on our calculation. The resonance formation, here
around $6$TeV, brings an enhancement factor of more than $4$
orders of magnitude. On the other hand departure from the full
one-loop calculation is of relevance only for $\mneuto$ masses
around $1$TeV. The insert in Fig.~\ref{fig.plateauM1japon} shows
in more detail the comparison of the full calculation compared to
the approximate result for the smaller higgsino LSP masses, well
before the resonance effects settle in. For $\mneuto > 600$GeV the
approximation is very good, only around $\mneuto \sim 200$GeV, the
full calculation captures the effect of other contributions like
those of Fig.~\ref{fig.feyn-gg}b,e). For this particular case it
looks like a good matching between the full one-loop result and
the non perturbative one should be made at around
$\mneuto=400$GeV. A possible strategy for choosing this matching
point would have to rely on the knowledge of both the full
one-loop result, the approximate one-loop result as given
Eq.~\ref{asymptotics-reduct} and the non-perturbative results
based on the fitting functions of Ref.~\cite{lspgg-nojiri}. This
would, of course,  only be carried out in the limit of almost pure
higgsino or wino. We would then have to compare the three results.
To revert to the non-perturbative regime means that the full
one-loop and the approximate one-loop agree fairly well and are
quite different from the non-perturbative regime. If, on the other
hand,  these two one-loop results differ sensibly this means that
one is not quite in the asymptotic region and that we might be
missing some one-loop contributions. If this is the case one
should also expect the higher order effects computed for the
threshold region to be small so that the non-perturbative result
and the approximate one-loop are very similar. Of course, as shown
in the example of Fig.~\ref{fig.plateauM1japon} these differences
in the higgsino region, compared to  taking the perturbative
parameterisation, are rather small compared to the uncertainty
that is inherent in the astrophysics part of the prediction of the
gamma ray line. For the wino, as we saw, the transition to the
asymptotic value is rather drastic especially for TeV LSP's,
therefore one should quickly capture  the non-perturbative regime.
Especially in this case one should also revert to a one-loop use
of the chargino-LSP mass difference.

\section{Conclusions}
There has been a flurry of activity in the last few years in the
search of dark matter with, among other strategies, many
experiments dedicated to the indirect searches of dark matter in
particular the gamma ray signal. The mono-energetic gamma ray line
signal constitutes a  clean signature. The improvement in coverage
and accuracy of the measurements should be matched by precise
theoretical calculations that should be publicly available through
general purpose codes. In this paper we have provided a new
calculation for \nngg and \nnzg for the annihilation of the
supersymmetric dark matter candidate and rederived as a bonus the
\nnglgl rate. These calculations have been made both for
small (zero) relative velocity of the neutralinos as adequate for
annihilation in the halo of our galaxy for example, but also for
velocities that would be needed for the contribution of these
channels in a precise derivation of the relic density. For \nngg
and \nnzg at $v\neq 0$ as would be needed for an improved relic
density prediction, these results are new. For \nngg
three\cite{lsptogg,lsptoggfernand} full one-loop calculations
performed for $v=0$ have already been performed. We have performed
a tuned comparison with the results of  {\tt
DarkSUSY}\cite{DarkSUSY} and {\tt PLATON}\cite{lsptoggfernand} and
have found perfect agreement. The calculation of
\nnzg is trickier and can not just be deduced from \nngg. Until
now there has been only one calculation\cite{lsptozg} of this
process. The latter has missed some contributions that may not
always be negligible. Comparisons of our results with the previous
ones without these contributions are quite good for scenarios
where the cross section is not small. In this paper we have not
made an attempt to fold in with the astrophysics part that
involves, for example, the halo profile but concentrated on the
particle physics part which must be an unambiguous prediction. To
pave the way for an implementation in {\tt
micrOMEGAs}\cite{micromegas-all} we felt it was important to
critically review the large mass higgsino and wino LSP scenarios
especially that the latter give large cross sections. As shown
in\cite{lspgg-nojiri} one needs to go beyond the one-loop
treatment in this regime. We have argued how one could match the
full one-loop treatment with the non-perturbative
result. \\

\noi Another important aspect of this paper is the way all these
calculations have been performed in a unified manner and the
techniques that we used. These processes are the first application
of a code for the calculation of one-loop processes in
supersymmetry both at the colliders and for astrophysics/relic
density calculations that require also a new way of dealing with
the reduction of the tensor integrals. The calculations are
performed with the help of an automatised code that allows gauge
parameter dependence checks to be performed. The use of a
generalised  non-linear gauge is crucial.

\vspace*{1cm}

\noi {\bf Acknowledgment}\\
We acknowledge discussions with our other colleagues of the {\tt
micrOMEGAs} Project in particular G.~B\'elanger who made some
early comparisons with {\tt DarkSUSY} and A.~Pukhov who made some
test runs with the $\gamma \gamma$ and $Z\gamma$ annihilations
routines within \muo. P.~Brun and S. Rosier-Lees also made some
test runs for the gamma ray predictions using a private version of
\muo$\;$ for indirect detection and we gratefully thank them here.
F.B. would like to thank the members of the Minami-Tateya group
for important discussions and a fruitful collaboration. He also
acknowledges discussions on the Gram determinant issue with
T.~Binoth, S.~Dittmaier, J.P.~Guillet and E.~Pilon. F.~Renard has
been, as always, of invaluable help. We also thank J.~Hisano and
M.~Nojiri for confirming that Eq.~61 of hep-ph/0412403 should be
squared and that the entry $i=0,j=1$ in Table~1 of that paper is
10 times smaller. P.~Gondolo has promptly provided us with the
results of {\tt DarkSUSY} reported in Table~1 and explained how to
enter the parameters within this code for our tuned comparison.
Fig.~\ref{fig.feyn-gg} and Fig.~\ref{fig.feyn-zg} are, in part,
generated with {\tt JaxoDraw}\cite{jaxodraw}. This work is
supported in part by GDRI-ACPP of the CNRS (France). The work of
A.S is supported by grants of the Russian Federal Agency of
Science NS-1685.2003.2 and RFBR 04-02-17448 and that of D.T is
supported, in part, by a postdoctoral grant from the Spanish
Ministerio de Educacion y Ciencia.

\renewcommand{\thesection}{\Alph{section}}
\setcounter{section}{0}
\renewcommand{\theequation}{\thesection . \arabic{equation}}
\setcounter{equation}{0}

\renewcommand{\thesection}{\Alph{section}}
\setcounter{section}{0}

\section{Appendix: Segmentation of loop integrals}

The tensor integral of rank $M$ corresponding to a $N$-point
graph, $\{M,N\}$, that we encounter in the general non-linear
gauge but with Feynman parameters $\xi=1$ are such that $M \leq
N$. For the evaluation of $2 \ra 2$ processes $N=4$ is the maximum
value and corresponds to the box. The general tensor integral
writes as

\beqn
\label{int-M-N} T_{\underbrace{\mu \nu \cdots \rho}_{M}}^{(N)}=\int
\frac{d^n l}{(2\pi)^n} \; \frac{l_\mu l_\nu \cdots l_\rho}{D_0 D_1
\cdots D_{N-1}}, \quad M \leq N,
\eeqn
\noindent where

\beqn
\label{Di} D_i=(l+s_i)^2-M_i^2, \quad s_i=\sum_{j=1}^{i} p_j, \quad
s_0=0.
\eeqn
\noindent $M_i$ are the internal masses, $p_i$ the incoming momenta
and $l$ the loop momentum. \\

The $N$-point  scalar integrals correspond to $M=0$. All higher
rank tensors for a $N$-point function, $M\geq 1$, can be deduced
recursively from the knowledge of the $N$-point (and lower) scalar
integrals. In {\tt Looptools}\cite{looptools} this is based on the
Passarino-Veltman algorithm\cite{PassarinoVeltman}. In {\tt
Grace-loop} the implementation is outlined in
Ref.~\cite{nlgfatpaper}. The tensor reduction involves solving,
recursively, a system of equations which explicitly requires the
evaluation of the Gram determinant: $Det G(p_1,p_2,p_3)=Det
G_{123}=Det p_ip_j$. For special kinematics the latter vanishes or
can get very small, leading to severe numerical instability. This
special kinematics for the general $2
\ra 2$ process one encounters in high energy occurs for
exceptional points in phase space, for instance in extremely
forward regions and most generally the weight of this contribution
may be dismissed. For the case at hand, when the two neutralinos
are at rest, or with extremely low relative velocity, the Gram
determinant vanishes for all points because the incoming
neutralinos have the same momentum and can not be considered
independent. This is exactly what occurs in our case. Indeed here,
the box diagrams for \nnzg have the Gram determinant

\beqn
Det G=M_{\neuto}^6 v^2
\frac{\sin^2\theta}{(1-v^2/4)^3} (1-z^2), \quad z^2=\frac{M_Z^2}{4
M_{\neuto}^2}(1-v^2/4),
\eeqn
$v$ is the relative velocity and $\theta$ is the scattering angle.
For \nngg, $z=0$. In our application the sub-determinant with the
incoming LSP neutralinos is responsible for the vanishing of the
box Gram determinant for all angles:

\beqn
Det G (p_1,p_2)=-M_{\neuto}^4 v^2
\frac{1}{(1-v^2/4)^2}.
\eeqn
This also means that the reduction of the tensor integrals for
triangles with the two LSP as external legs, needs special
treatment. Such triangles are of the type Fig.~1.e) (but not
Fig.1~d)).

We therefore need to implement a routine for cases when the
determinant vanishes due to the fact that the momenta $s_i, i\neq
0$ are not independent. There are a few ways of dealing with the
tensor integrals when the Gram determinant is exactly
zero\cite{thomasfatpaper,robin-reduc}. Sometimes, the problem is
even avoided by the grouping of terms so that this spurious
inverse determinant cancels out. Our aim was to find, at least for
$2 \ra 2$ process, an efficient method that can, not only be
easily automated, but also calls most of the existing routines
that are present for a general purpose algorithm designed for non
zero Gram determinants. Take the box for example. Observe that (in
any $n$ dimension), in most generality, we may write for any given
pair of constants $\alpha,\beta$

\beqn
\label{eq:segment} \frac{1}{D_0 D_1 D_2 D_3}&=& \left(\frac{1}{D_0
D_1 D_2} -\alpha \frac{1}{D_0 D_2 D_3}  -\beta \frac{1}{D_0 D_1 D_3}
+(\alpha+\beta
-1)\frac{1}{D_1 D_2 D_3} \right) \times \nonumber \\
& & \frac{1}{A+2 l.(s_3-\alpha s_1 -\beta s_2)} \nonumber \\
A&=&(s_3^2-M_3^2)-\alpha (s_1^2-M_1^2)-\beta (s_2^2-M_2^2)-(\alpha
+\beta-1) M_0^2.
\eeqn

Obviously if $s_3=\alpha s_1+\beta s_2$ and hence the momenta are
linearly dependent, the box splits into a sum of triangles. We
will refer to this as segmentation. This segmentation is
independent of the tensor structure. This means that the reduction
of the tensor box with zero Gram determinant amounts to a tensor
reduction for triangle with a non-zero Gram determinant for which
one uses the usual procedure and hence uses the general library.
Observe that if $\alpha=0$ or $\beta=0$, there are three segments
instead of four. The missing segment, triangle integral, does in
fact have a zero Gram determinant. Therefore when one approaches
the zero of the Gram determinant of the box in these specific
cases, $\alpha \sim 0$ for example, $\alpha$ will be numerically
very small but non zero. A numerical instability could still
develop due now to the Gram determinant of the associated
triangle. These ``algebraic" zeros could be missed at the
numerical level. This can again lead to (less severe) numerical
instabilities due to the reduction of the associated tensor. In
this case, these triangles are segmented even further into
two-point functions, following the same recipe. This way their
contribution is  negligible even at the numerical, automatic,
level. In any case, since we also encounter (tensor) triangle
diagrams (Fig.~1.e)) that have a vanishing Gram determinant, we
have also included a segmentation that also applies to the tensor
triangles using the same trick as for the boxes.

There is an important observation to make. The segmentation of a
tensor of rank $M$ for the $N$-point function, $\{M,N\}$, amounts
to applying the tensor reduction on $\{M,N-1\}$. If $M=N$, after
segmentation one would need a library for $\{N,N-1\}$. These are
not supplied by default in the general libraries. These libraries
would then need to be extended. Reduction of $N-1$-point function
tensor integrals are much more compact and easier than for
$N$-point functions. This said for the case at hand, and for that
matter any relic density calculation where the LSP is a
neutralino, these highest rank tensors are not needed. It is easy
to show that the highest rank tensors  for $2\ra 2$ only occurs
when all the external particles are bosons. In our case, for the
box, one has $M_{{\rm max}}=3$. For $\neuto \neuto \ra f
\bar f$, $M_{{\rm max}}=2$.

The choice of the momenta circulating in the loop, $s_1,s_2,s_3$,
is not unique and depends on the particular graph. If $Det
G(s_1,s_2,s_3)=0$ we first form all three sub-determinants $Det
G(s_i,s_j)$ and take the couple $s_i,s_j$ that corresponds to $Max
\; |Det(s_i,s_j)|$. Then the third $s_k$ is distributed in the basis
$s_i,s_j$ and the corresponding $\alpha$ and $\beta$ are read. In
fact, suppose $Det G(s_1,s_2) \neq 0$, then it is revealing to
{\em always} write

\beqn
s_3&=&\alpha s_1 +\beta s_2 + \varepsilon_T \quad {\rm with} \nonumber \\
s_1.\varepsilon_T&=&s_2.\varepsilon_T=0 \quad {\rm meaning} \nonumber \\
\varepsilon_{T,\mu}&=&\epsilon_{\mu \alpha\beta \delta} s_1^\alpha
s_2^\beta t^\delta.
\eeqn

\noi $\varepsilon_T$ is a vector that is orthogonal to both $s_1$ and
$s_2$ that is easily reconstructed once $\alpha$ and $\beta$ are

\beqn
\alpha=\frac{s_2^2 s_3.s_1 -s_1.s_2 s_2.s_3}{Det G(s_1,s_2)}, \quad
\beta=\alpha (s_1\leftrightarrow s_2).
\eeqn

\noi This construction makes it clear that

\beqn
Det G(s_1,s_2,s_3)=\varepsilon_T^2  Det G (s_1,s_2).
\eeqn

\noi This shows, in a most transparent manner, that the determinant
vanishes when $\varepsilon_T^2=0$. This can occur when the
components of this vector vanish, $\varepsilon_T^\mu \ra 0$, and
therefore $s_3$ is not an independent vector as the case of this
paper for $v=0$. It also occurs, a point which is often
overlooked, when $\varepsilon_T$ is light-like. However the
orthogonality constraint means that the other vectors $s_1,s_2$
are space-like. Therefore this case does not occur for real
particles that are time-like, and hence does not occur for our
$2\ra 2$ process. It will be shown, in a separate publication,
that when $\varepsilon_T$ is light-like a segmentation is still
possible. The algorithm can also be improved by expanding around
$Det G(s_1,s_2,s_3)=0$. We will come back to the details of this
issue in a future publication. Note that for $v=0.5$ and for all
three processes studied in this paper the standard reduction
formalism, without segmentation, was used.

\end{document}